\documentclass{ws-p8-50x6-00}
\usepackage{floatflt}
\newlength{\wi}   \wi 8cm
\newlength{\fwi} \fwi 0.95\wi
\begin{document}
\title{Quark and gluon jet properties at LEP}

\author{Martin Siebel}

\address{Dep. of Physics, University of Wuppertal, Gau\ss stra\ss e 20,
D-42097 Wuppertal,
Germany\\E-mail: Martin.Siebel@cern.ch}

\maketitle
\abstracts{
The study of the differences of the fragmentation of quarks and gluons
to jets of hadrons gives insight into the fundamental structure of
QCD. Results from different approaches to properties of quarks and
gluons are shown. The colour factor ratio $C_A/C_F$ 
is measured in agreement with the QCD prediction.
Identified particles in
quark and gluon jets are investigated, revealing no overproduction
of isoscalar $\eta^0$ and $\phi(1020)$ in gluon jets, but an excess of protons.
}
\section{Introduction}
In QCD there are three fundamental vertices representing the
three and the four gluon coupling and the coupling of a fermionic
current to a gluon.
Neglecting the four gluon coupling which is of ${\cal O} (\alpha_s^2)$
there are three fundamental processes described by these vertices:
The radiation of a gluon by a quark ($q\to qg$) or by another gluon
($g \to gg$) and the
splitting of a gluon into a quark-antiquark pair ($g \to q\bar q$). 
The probability of
these processes is given by the splitting kernels which take the
kinematics into account and are proportional to the colour factors
$C_F$, $C_A$ and $T_F$ respectively. The colour factors are the 
structure constants
of the $SU(3)$ colour symmetry group of QCD,
 vividly speaking, they
take care of the bookkeeping of undetermined colour flows
in these processes. 
%Additionally the splitting kernel for the process
%$g\to q\bar q$ is proportional to $n_f$, the number of active
%quark flavours, to consider the undetermined flavour of the outgoing
%$q\bar q$ pair. 
As the main contribution to particle production in
hadronic events is due to the radiation of gluons, it is to be 
expected that the differences of the properties of jets initiated by 
quarks to those of jets initiated by gluons are in terms of the ratio
$C_A/C_F$. The assumption of {\em local parton-hadron duality} (LPHD) gives
rise to the hope that these perturbatively calculated differences 
can be observed
in the hadronic final state. 
\vspace{-.3cm}
\section{Experimental access}
\vspace{-.1cm}
To get information about gluons, events with three jets are used, as
they contain exactly one gluon jet. 
The jets of a hadronic three jet event are
numbered according to their energy with jet 1 being the most energetic
one while angles between jets are numbered according to the jet
opposite of the angle.
There are three different ways of getting information about quarks and
gluons out of hadronic three jet events presented here.
In
the {\em jet analysis}\cite{multi,scaling} single jets taken from
three jet events are
looked at. 
Gluon jets are identified using an anti-tagging technique in events
with primary $b$-quarks.
Information about light quark jets is obtained by subtracting the 
properties of gluon jets from a sample of
unidentified jets.
Approximately 142.000 identified gluon jets enter this analysis. 
As the dynamical scale for a single jet the so called 
{\em hardness scale}
$\kappa_h=E_{jet}\cdot\sin\theta_{\min}$ is used, where  
$\theta_{\min}$ represents the smaller of the two angles adjacent to a
jet. 
Although this is a problem which, in principle, involves two scales, 
$\kappa_H$
proves to be an appropriate choice to reduce this to a one scale problem.
The jets entering this analysis
cover the 
dynamical range of approximately $\kappa_h=6GeV
\dots 29GeV$.
In the {\em hemisphere analysis}\cite{ojets} events are selected with
a hard gluon
initiating the most energetic jet. 
The gluon jet is defined by all particles in the hemisphere of the
leading jet.
This type of event requires both subleading jets to be tagged as
$b$-quark jets successfully. 
Light quark jets to compare with are defined by a whole
hemisphere of an event.
This procedure gives 439 gluon jets at an energy of $40.1GeV$
and quark jets at $45.6GeV$. Corrections for the difference in energy
are applied to the quark jets.
In the {\em symmetric event analysis}\cite{multi,kegel} complete three
jet events are investigated.
Only events with $\theta_2$ being equal to $\theta_3$ within small tolerances
($\pm 1.25^\circ$) are used so the whole event can be described by
giving only $\theta_1$. No assignment of particles to jets and no jet
tagging is needed.
A total of approximately $162.000$ symmetric events enter this
analysis.  
\vspace{-.3cm}
\section{Results}
\vspace{-.1cm}
\wi 0.4\textwidth
\begin{figure}[tb]
  \begin{minipage}[t]{\wi}
    \begin{center}
       \epsfxsize=11pc % will enlarge or reduce the postscript figures based on the xsize
       \epsfbox{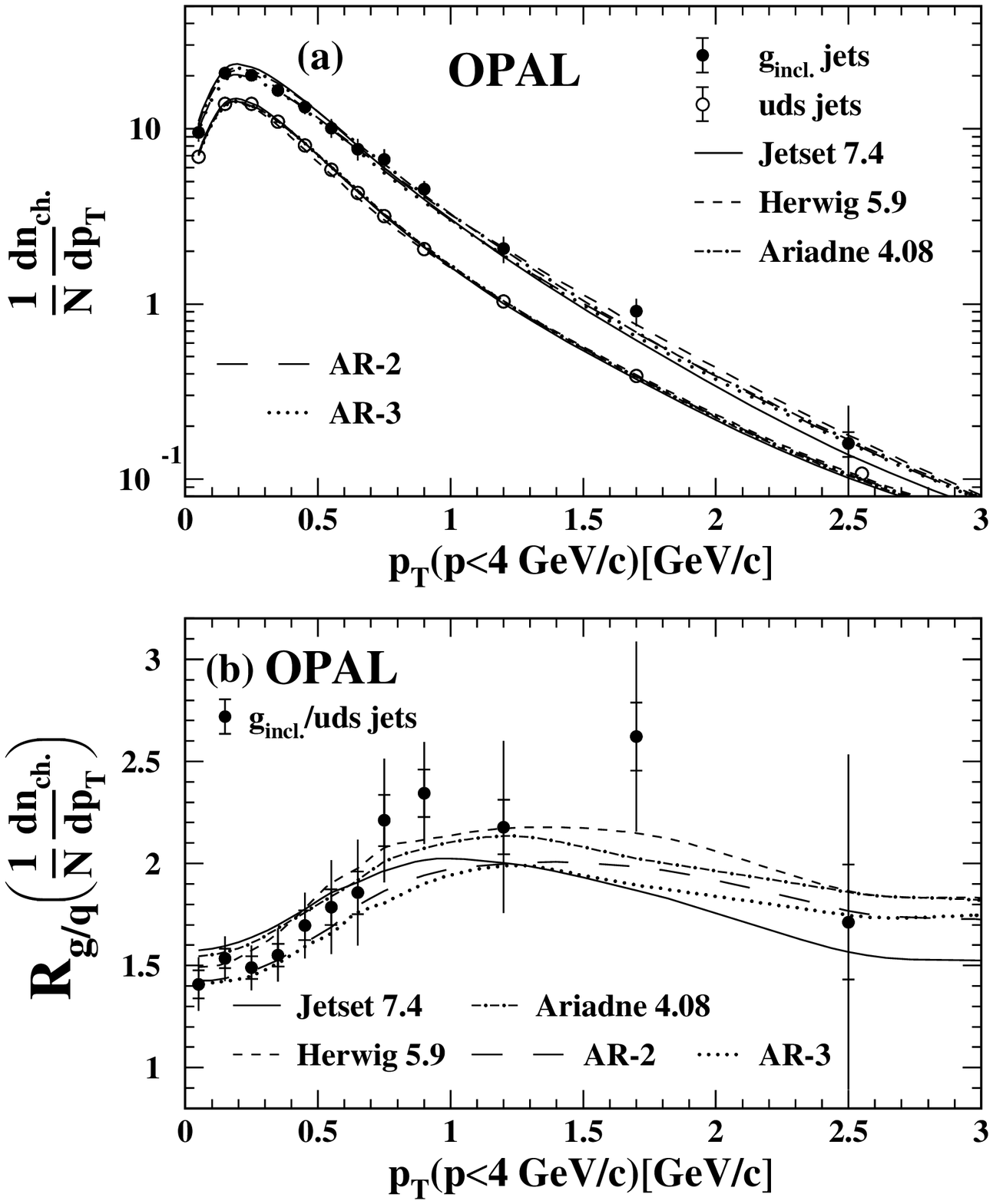} % postscript image file name
\caption{$p_T$-distributions (upper plot) and
the ratio gluon/quark of the distributions (lower plot)\label{fig:opalpt}}
     \end{center}
   \end{minipage}
   \hfill
   \wi 0.5\textwidth
   \begin{minipage}[t]{\wi}
     \begin{center}
       \epsfxsize=13pc % will enlarge or reduce the postscript figures based on the xsize
       \epsfbox{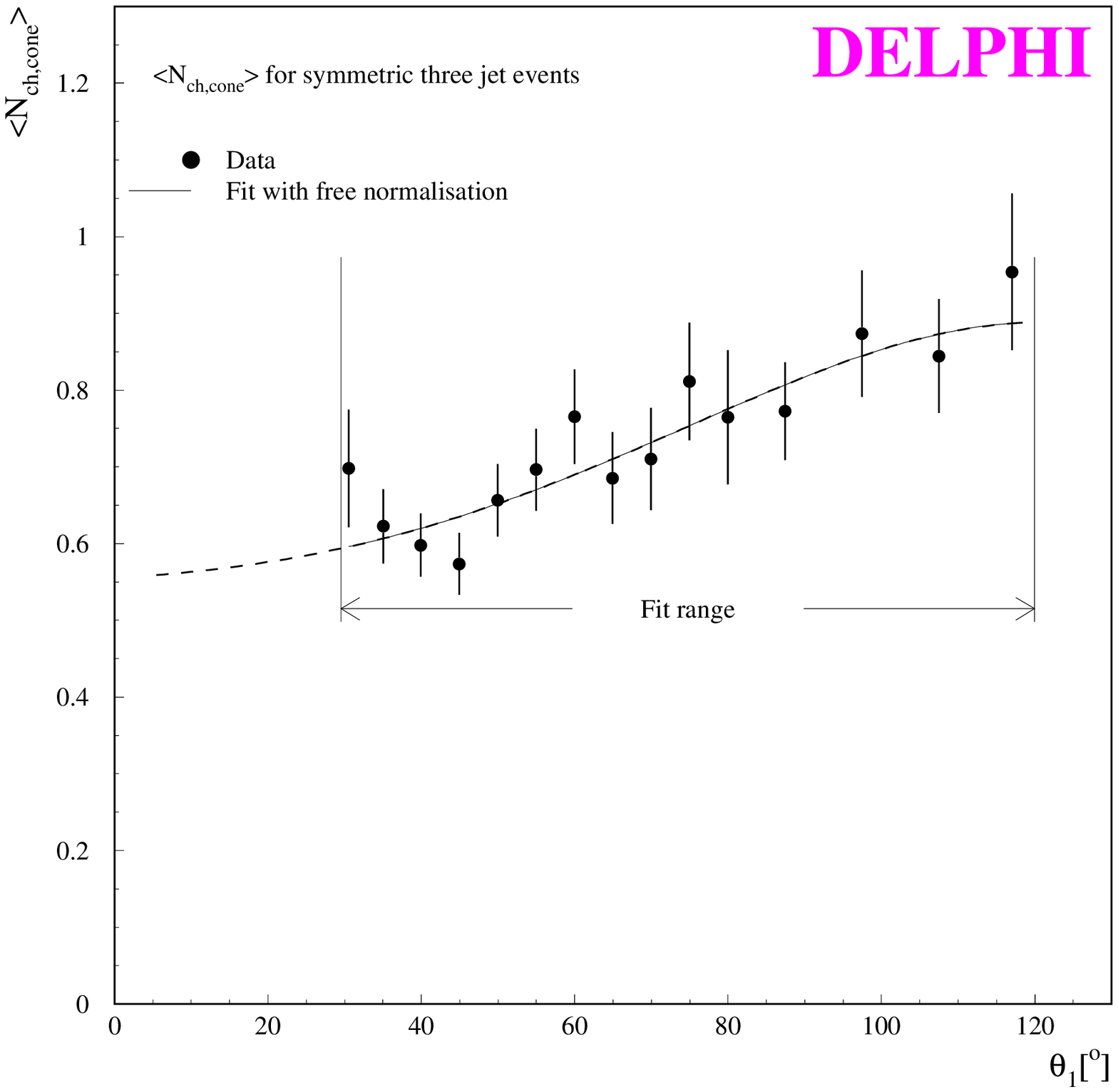} % postscript image file name
       \caption{The multiplicity within a $30^\circ$-cone perpendicular to the
        event plane as function of the event topology \label{fig:kegel}}
     \end{center}
   \end{minipage}
\vspace{-0.5cm}
\end{figure}
\wi 0.5\textwidth
\fwi 0.98\wi
\begin{floatingfigure}[hbt]{\wi}
\flushleft
\includegraphics[width=\fwi]{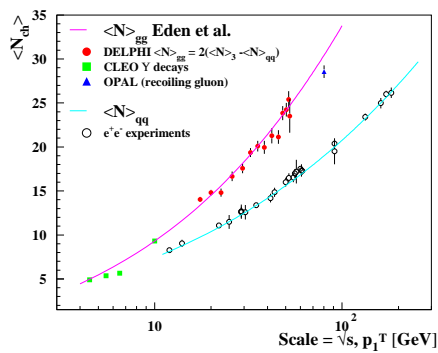}
\caption{The multiplicity of events at different energies (lower
curve) 
leads to a
prediction of the gluon contribution to multiplicity in three jet
events (upper curve).\label{fig:eden}}
\end{floatingfigure}
As stated above, the gluon bremsstrahlung is expected to be $C_A/C_F$
times higher in gluon than in quark jets.
This contrasts to the
experimental observation of the ratio $r_{ch}$ of the charged
multiplicity in gluon
jets over quark jets to be below $1.5$ as found over the
whole scale range of the {\em jet analysis}\cite{multi} and $1.514\pm
0.019 \pm 0.034$ for the {\em hemisphere analysis}\cite{ojets}.
This
difference to the na$\ddot{\mathrm\i}$ve expectation can be explained
by higher order corrections and effects of finite energy. Soft
partons emitted at large $p_T$ cannot resolve single
partons\par\noindent
within the
shower-evolution of a jet, so 
the effective colour charge emitting these particles is the one
of the primary parton.
Figure \ref{fig:opalpt} shows the $p_T$ distribution of
soft particles $(|{\bf p}|<4 GeV/c)$ in quark and gluon jets and the ratio of
both in the lower plot as taken from the {\em hemisphere analysis}. 
For higher $p_T$ the ratio is in the expected
range of $C_A/C_F=2.25$. Monte-Carlo studies confirmed that this ratio in
the range of $0.8GeV/c<p_T<3GeV/c$ reflects the colour factor ratio,
yielding $r_{ch}(|{\bf p}| < 4 GeV/c, 0.8GeV/c<p_T<3GeV/c)= 2.29
\pm 0.09 \pm 0.15$ in good agreement with the QCD expectation for 
$C_A/C_F$\cite{ojets}.
The same idea is used in figure \ref{fig:kegel} from the {\em event
analysis}.  
This plot shows the multiplicity within a cone with fixed
opening angle of $30^\circ$ perpendicular to the event plane. The
abscissa is the angle $\theta_1$ between jet 2 and 3. The plotted line
is derived from a prediction by V.Khoze, W.Ochs and S.Lupia\cite{kol} and
parametrised by
\begin{equation}
N_\perp^{q\bar qg} \propto 2+\cos\frac{\theta_1}{2}-\cos\theta_1-
\frac{1}{N_C^2}\left(1+\cos\frac{\theta_1}{2}\right)
\end{equation}
with only the normalisation left free\cite{kegel}. 
The prediction describes the
data well, giving evidence of coherent radiation within a three jet
event. 
\begin{figure}[b]
  \begin{minipage}[b]{\wi}
    \begin{center}
       \epsfxsize=14pc % will enlarge or reduce the postscript figures based on the xsize
       \epsfbox{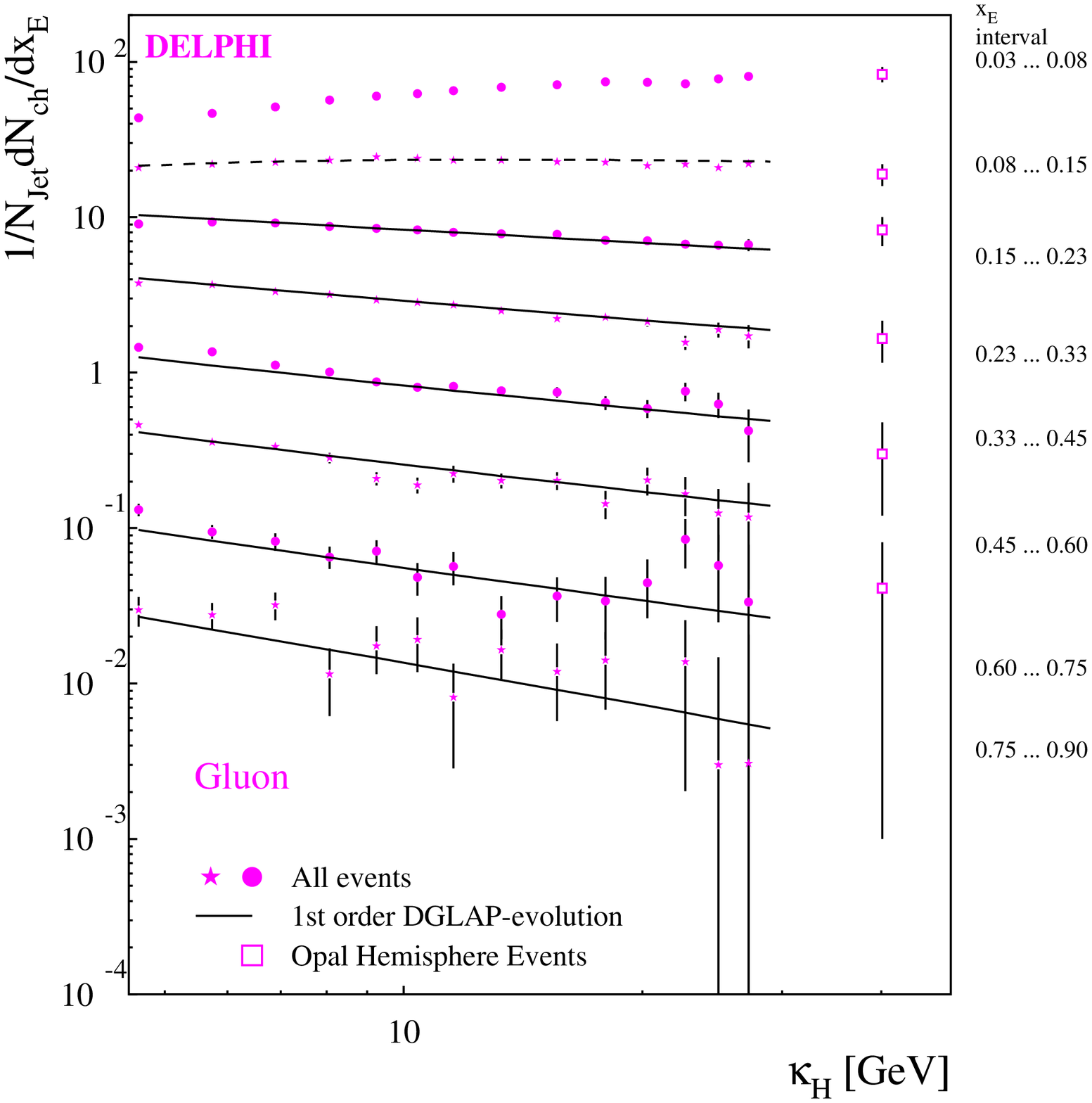} % postscript image file name
\caption{The fragmentation function of the gluon as a function of
$\kappa_H$. 
\label{fig:scaling}}
     \end{center}
   \end{minipage}
   \hfill
   \wi 0.45\textwidth
   \begin{minipage}[b]{\wi}
     \begin{center}
       \epsfxsize=13pc % will enlarge or reduce the postscript figures based on the xsize
       \epsfbox{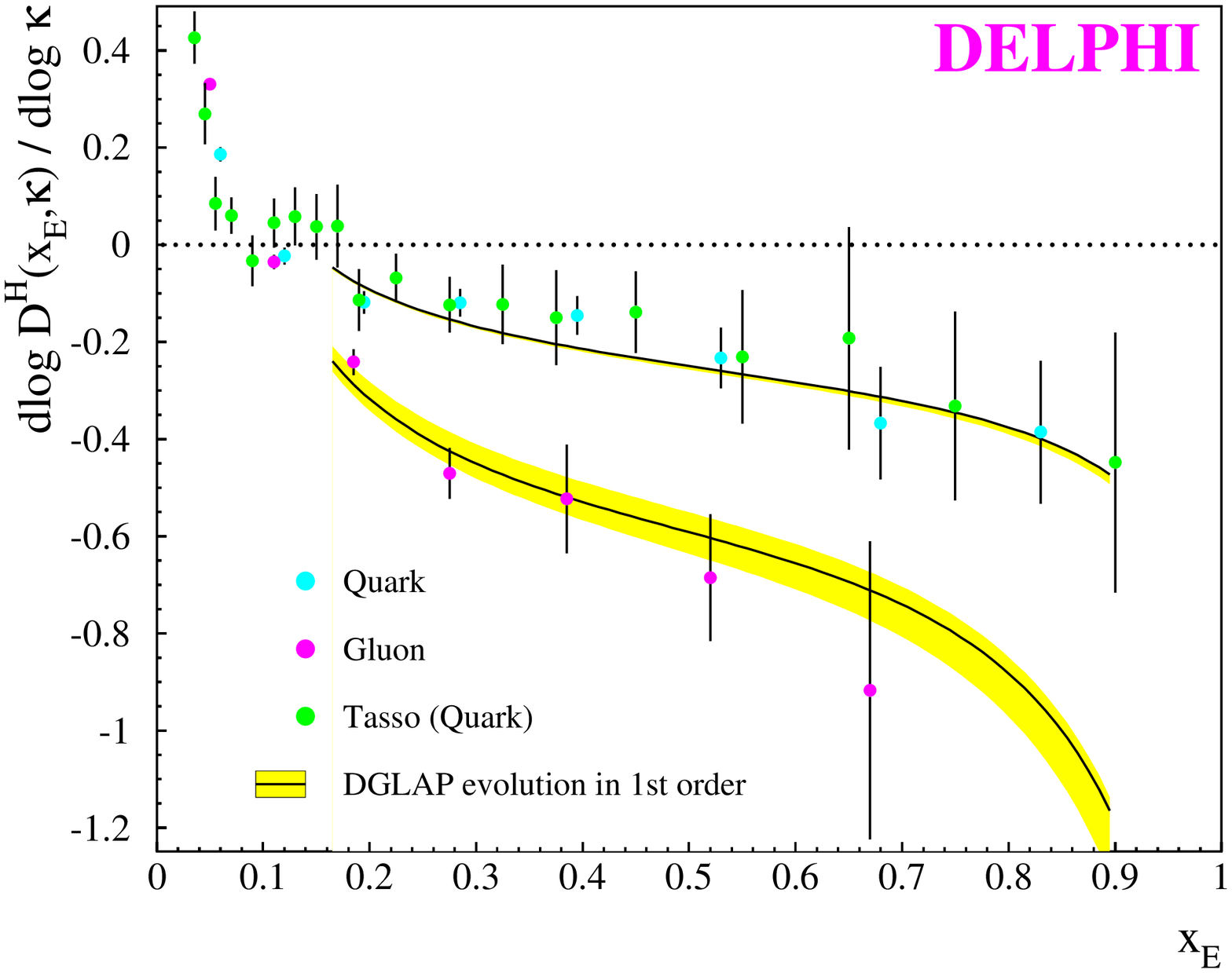} % postscript image file name
\caption{Strength of the scaling violation in quark and gluon jets as
a function of $x_E$. 
\label{fig:violation}}
     \end{center}
   \end{minipage}
\vspace{-0.5cm}
\end{figure}
This provides a direct test of a perturbative calculation
independent of fragmentation models. 
Both approaches are complementary, as in the {\em hemisphere analysis}
the topology and therefore the colour charge separation is fixed while
partons of different wavelengths and resolution capabilities have been
investigated, while in the {\em symmetric event analysis} the cone and
therefore the phase space of the partons remains fixed and the
separation of the colour charges is varied with the topology.
\par
For the multiplicity of a hadronic three jet event there is a
MLLA-prediction\cite{eden} of the simple form
\begin{equation}
N_{q\bar qg}(\theta_1)=N_{q \bar q}(2E^*,p_1^t )+\frac{1}{2}N_{gg}(p_1^t)
\label{eq:eden}
\end{equation}
where coherence effects are taken into account by the used scales
$2E^*$ and $p_1^t$. As in the {\em symmetric event analysis} the event
is described by only giving $\theta_1$, the two scales $2E^*$ and 
$p_1^t$ can be
expressed as functions of this angle. The contributions $N_{q \bar q}$
and $N_{gg}$ are correlated by 
a differential equation
including the colour factor ratio with corrections. Taking the
parametrisation of $N_{q\bar q}(\sqrt{s})$ from events at
various beam energies and fixing the only free parameter left by a
measurement of the $gg$-multiplicity by {\sc Cleo}\cite{cleo}, one gets an
absolute prediction of the topology dependence of the three jet
multiplicity. 
This prediction is found to be in very good agreement 
with the data
from the {\em symmetric event analysis} down to an opening angle of
$\theta_1\simeq30^\circ$. The lower curve in figure \ref{fig:eden} shows
the parametrisation $N_{q\bar q}(\sqrt{s})$ of the data.
The upper curve is $N_{gg}$ derived
from $N_{q\bar q}$, the corresponding data points from the {\sc
Delphi} {\em symmetric event analysis} are calculated from three jet event 
multiplicities by subtracting $N_{q\bar q}$. The $N_{gg}$ data
point at $10GeV$ from {\sc Cleo} has been used to fix the prediction
while the point at
$40GeV$ is taken from {\sc Opal}'s {\em hemisphere
analysis}\cite{ojets}. 
The lowest
three points from the {\sc Cleo} analysis are without systematic
errors\cite{cleo}. The overall agreement between data and prediction
is good. The result for
$C_A/C_F$ gained by the {\em symmetric event analysis} is 
$C_A/C_F=2.246 \pm 0.062 \pm 0.080 \pm 0.095_{theo}$ where an ansatz
has been used which takes care of non-perturbative effects by an added
offset\cite{multi}. Numeric results from a fit with equation
\ref{eq:eden} are to be expected soon.\par
While the multiplicity is dominated by soft particles, the study of
scaling violation in quark and gluon jets offers the possibility to
obtain information from the other end of the spectrum. 
The energy dependence of the fragmentation functions is described by the 
DGLAP-equations. As the splitting
kernels occur in this set of integro-differential equations, again the
colour factor ratio $C_A/C_F$ is expected to be found in the
comparison of gluon to quark jets. In the {\em jet analysis} the
fragmentation functions of quarks and gluons have been 
investigated\cite{scaling}. Figure \ref{fig:scaling} shows the
fragmentation function of the gluon as a function of $\kappa_H$ and
$x_E$ as a parameter.
$\kappa_H$
is used as the dynamical scale
as we are dealing with single jets. 
The curves are the result of an evolution of the
DGLAP-equations to first order. The description of the data is very good
and extrapolates well to the data taken from the {\em hemisphere analysis}
at the rightmost\cite{ojets}. Fitting a power-law ansatz through the 
measured fragmentation
functions yields the result shown in figure
\ref{fig:violation} where the double logarithmic slopes found in gluon and
quark jets are plotted. Obviously the scale dependence is more pronounced in
gluon than in quark jets. Again the lines represent the results of the
evolution of the DGLAP equations. A fit with this evolution
yields for the colour factor ratio $C_A/C_F=2.25\pm
0.09\pm 0.06\pm 0.11$ where the last error reflects the dependence on
the jet finding algorithm and choice of scale. 
This is in excellent agreement with the QCD 
prediction.
%
%
%
%
%
%\section{Identified particles}
%
\wi 0.5\textwidth
\fwi 0.98\wi
\begin{floatingfigure}[t]{\wi}
\flushleft
\includegraphics[width=\fwi]{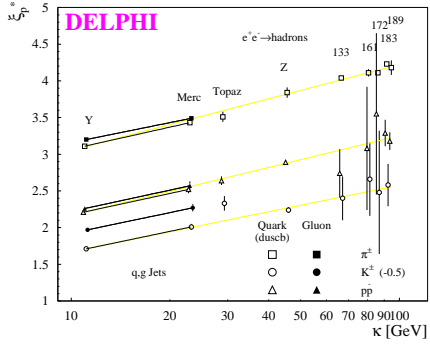}
\caption{Position of the maximum of the $\xi_p$ distribution. The
values for $\xi$ are obtained by using the decimal logarithm.
\label{fig:xistar}}
\end{floatingfigure}
Also identified particles in quark and gluon jets have been studied,
especially $\pi^\pm$, $K^\pm$, $p\bar p$,
$\pi^0$ and $\eta$ as well as the resonances 
$K^\ast(892)$ and $\phi(1020)$\cite{idpart,opart}. 
These analyses both have dealt with single
jets similar to the {\em jet analysis} described above. The
investigation of the multiplicity of these particles in quark in gluon
jets is of interest especially for the $\eta$ and $\phi(1020)$, as
these mesons are isospin singletts as the gluons are. Despite this
community, no overproduction of these two particles has been found in
gluon jets. 
The multiplicity of $\eta$ particles has been studied at three
different values for $\kappa_H$ between $7GeV$ and $25GeV$. The
multiplicity found in quark and in gluon jets as well as the ratio of
both always agreed with the expectations from charged particles\cite{opart}.
For the resonances 
$K^\ast(892)$ and $\phi(1020)$ the normalised ratio 
$R_X=[n_g(X)/n_q(X)]/[n_g(ch)/n_q(ch)]$ with $n_{q,g}(X)$ being the
multiplicity of particle $X$ in quark or gluon jets and $ch$ denoting
all charged particles has been investigated. The results obtained 
are $R_{K^\ast}=1.7\pm
0.5$ and $R_\phi=0.7 \pm 0.3$ revealing especially no
excess of $\phi$ in gluon jets\cite{idpart}.
While also pion and kaon production
have been found in agreement with the results for all charged
particles, a significant excess of protons in gluon jets has been
found with $R_p=1.205 \pm 0.041 \pm 0.02$\cite{idpart}.
Also the momentum spectra and $\xi_p$ distributions\par\noindent of these particles
have been studied showing a good agreement with the predictions of
{\sc Jetset} and {\sc Ariadne} and more pronounced deviations from
{\sc Herwig} predictions. The position of the maximum of the $\xi_p$ 
distributions, $\xi_p^\ast$, is shown in figure \ref{fig:xistar} for protons,
pions and kaons. The data
points denoted with Y and MERC are taken from a jet analysis in which
only two symmetric event topologies where used\cite{idpart}. 
The behaviour of the quark $\xi_p^\ast$ values from
this jet analysis extrapolates well to the measurements from 
events at higher energies. For pions and protons the $\xi_p^\ast$ values
found in quark and gluon jets are in good agreement, while for kaons
the $\xi_p^\ast$ values found in gluon jets are significantly higher than
in quark jets. This observation does not agree with the assumptions of
LPHD.
\vspace{-.3cm}
\section{Acknowledgements}
\vspace{-.1cm}
I would like to thank 
J.Drees to whom I owe the opportunity to work in this fascinating
field and 
K.Hamacher, O.Klapp and P.Langefeld for the very
pleasant time working on this subject. I also would like to express my
gratitude to the organising committee for their effort in making the
ISMD'99 the delightful experience it has been.

\end{document}